\documentclass[reprint,aps,prl,nofootinbib,superscriptaddress]{revtex4-2}

\usepackage{comment}
\usepackage{aas_macros}
\usepackage{booktabs,chemformula}
\newcommand{\dd}{\mathrm{d}}

\newcommand{\uf}{\mathrm{f}}

\newcommand{\uA}{\mathrm{A}}
\newcommand{\uF}{\mathrm{F}}
\newcommand{\ub}{\mathrm{b}}

\newcommand{\Msun}{\mathrm{M}_{\odot}}
\newcommand{\Mchirp}{\mathcal{M}}
\newcommand{\CA}{C_{\mathrm{A}}}
\newcommand{\CF}{C_{\mathrm{F}}}

\newcommand{\zhat}{\mathbf{\hat{z}}}

\newcommand{\Dlb}{D_{\lambdabar}}
\usepackage{tikz}
\usepackage[compat=1.1.0]{tikz-feynman}
\usetikzlibrary{arrows.meta,decorations.pathmorphing}
\newcommand{\black}[1]{\textcolor{black}{#1}}

\usepackage{graphicx}
\usepackage{dcolumn}
\usepackage{bm}

\usepackage{ragged2e}
\usepackage{subcaption}
\makeatletter
\long\def\@makecaption#1#2{%
  \vskip\abovecaptionskip
  \centering 
  \parbox{\linewidth}{\justifying\small #1. #2\par}%
  \vskip\belowcaptionskip}
\makeatother

\usepackage[normalem]{ulem}

\usepackage[pdfnewwindow=true,      
   colorlinks=true,    
  linkcolor=blue,     
 citecolor=blue,     
filecolor=blue,  
urlcolor=blue,      
final=true,
]{hyperref}

\begin{document}
\setlength{\abovedisplayskip}{7pt}
\setlength{\belowdisplayskip}{7pt}

\preprint{APS/123-QED}

\title{Resonant Locking Between Binary Systems Induced by Gravitational Waves}

\author{Charlie Sharpe}
\affiliation{Rudolf Peierls Centre for Theoretical Physics, University of Oxford, Clarendon Laboratory, \\ Parks Road, Oxford OX1 3PU, United Kingdom \vspace{0.5em}}

\author{Yonadav Barry Ginat}
\affiliation{Rudolf Peierls Centre for Theoretical Physics, University of Oxford, Clarendon Laboratory, \\ Parks Road, Oxford OX1 3PU, United Kingdom \vspace{0.5em}}

\author{Zeyuan Xuan}
\affiliation{ Department of Physics and Astronomy, UCLA,\\ Los Angeles, CA 90095, USA\vspace{0.5em}}
\affiliation{Mani L. Bhaumik Institute for Theoretical Physics, UCLA, \\Los Angeles, CA 90095\vspace{0.5em}}

\author{Bence Kocsis}
\affiliation{Rudolf Peierls Centre for Theoretical Physics, University of Oxford, Clarendon Laboratory, \\ Parks Road, Oxford OX1 3PU, United Kingdom \vspace{0.5em}}
\affiliation{St Hugh's College, University of Oxford \\ St Margaret's Rd, Oxford OX2 6LE, United Kingdom \vspace{0.5em}}

\date{\today}

\begin{abstract}
The interaction of gravitational waves (GWs) with matter is thought to be typically negligible in the Universe. 
We identify an exception in the case of resonant interactions, where GWs emitted by a background binary system, such as an inspiraling supermassive black hole (SMBH) binary, cause a resonant response in a stellar-mass foreground binary and the frequencies of the two systems become, and remain, synchronized. \black{We point out that this previously unexplored dynamical phenomenon is not only possible, but can lead to $\mathcal{O}(30)$ binary systems becoming resonantly locked in the host galaxy of merging SMBHs of mass $10^{8.5-11}\Msun$, each of which has a significantly reduced merger time.
We predict $\mathcal{O}(10^{10})$ binary systems have been locked in the Universe's history. 
Resonant locking could be detected through anomalous inspiral of binary systems.}
\end{abstract}

\maketitle

\paragraph{Introduction---} The interaction of gravitational waves (GWs) with matter has been of interest since its prediction by \citet{Einstein_1916}. Bondi \citep{Bondi_1957, Bondi_1959} showed that GWs carry energy, and some of this energy can be dissipated in matter that the GWs pass through. Such dissipation can occur in an expanding Universe \citep{Hawking_1966, Hawking_1968}, a viscous fluid \citep{Esposito_1971}, radiation in the early Universe \citep{Papadopoulos_2002,Weinberg_2004}, neutrinos and dark matter \citep{Weinberg2004,Flauger_Weinberg2018},  galaxies and clusters \citep{Rees_1971,Futamase_1979,Loeb2020}, accretion disks \citep{Kocsis_2008, Li_2012}, oscillating stars \cite{Li_2012, Kojima_2005, McKernan_2014,Boyanov+2024}, binary systems \citep{Mashhoon_1978, Turner_1979, Bertotti_1983, Chicone_2000, Hui_2013, Iorio_2014, Annulli_2018, Rozneretal2020, Desjacquesetal2020, Blas_2022a, Kuntz_2023}, planets \citep{Bertotti_1973, Mashhoon_1981, Ismaiel_2011}, spacecraft \citep{ANDERSON_1971, Press_1972, Estabrook_1975, Rudenko_1975, Thorne_1976, Anderson_1976, Hellings_1978, Nelson_1982, Blas_2022b}, or crystals in laboratories \citep{Weber_1960}. \black{Gravitational waves act as periodic perturbations on astrophysical systems and, while the associated dissipation is ordinarily negligible away from resonance \citep{Mashhoon_1978, Turner_1979, Loeb2020}, the cumulative effect in resonance can dramatically enhance the response \citep{Chicone_1997a, Chicone_1997b, Chicone_2000, McKernan_2014, Rozneretal2020, Desjacquesetal2020}.}

Since it is known that Newtonian resonant perturbations commonly produce resonantly locked configurations for multi-planetary systems or moons in mean motion resonance, we are motivated to investigate if a similar capture of astrophysical systems into resonance is possible at a much larger distance from the source -- even in the wave zone, defined as the region outside of one reduced GW wavelength from the GW source. In particular, inspiraling supermassive black hole binaries (SMBHBs) are expected to form in most galaxies following galaxy collisions \citep{Volonteri_2003} whose GWs sweep through periods from thousands of years to hours, or even minutes, until they merge. Almost all binaries in the host galaxy will inevitably be in resonance with these GWs at some point during the SMBHB inspiral. 

In this paper, we investigate the phenomenon of a stellar-mass astrophysical binary system (``foreground, FG, binary'') undergoing resonant locking (or frequency synchronization) with the GWs emitted by an inspiraling SMBHB (``background, BG, binary'').
\color{black} 
See Figure~\ref{fig:systematic_setup} in the Appendix for an illustration of the set-up. We adopt geometrized units: $G = c = 1$.

\paragraph{Equation of Motion---} Table \ref{tab:Parameters} defines the binary parameters used in this work. Since the BG binary is much more massive than the FG binary, we neglect feedback effects. Subscripts `f' and `b' denote quantities of the FG and BG binaries, respectively. To leading order in $v_\uf/c$, the equation for geodesic deviation gives the relative force $F^i$ between the two FG binary components \citep[][Eq.~(8.43)]{Misner_2017}:
\begin{align}
    \frac{F^i}{\mu_\uf} = \ddot{r}^i_{\mathrm{f}} + \frac{M_\uf}{r^3_{\mathrm{f}}} r^i_{\mathrm{f}} = R^{0i0j} r_{\uf,j},
\end{align}
where dots ($\dot{}$) represent time derivatives, commas do \textit{not} refer to partial differentiation, and $R$ is the Riemann tensor due to the BG binary. The lowest-order (quadrupole) contribution to this force%
\footnote{\black{Higher moments at fixed frequency are suppressed by powers of $v_\ub/c$.}}
is \citep{Kuntz_2023}
\begin{align}
    \begin{split}
        &\frac{F^h}{\mu_\uf} = \frac{1}{D}r_{\uf,i} A_{1}^{ih} + \frac{1}{2D^3}r_{\uf,i} \left(D^i D_k A_{2}^{kh} + D^h D_k A_{2}^{ki} \right) \\
        &- \frac{1}{6D^3}r_\uf^h D_j D_k (2A_{2}^{jk} - A_{3}^{jk}) + \frac{1}{2D^5}r_{\uf,i} D^i D^h D_k D_l A_{3}^{kl},
    \end{split} \label{eq:Full force}
\end{align}
where repeated indices are summed over, $D^i$ is the vector pointing to the FG binary from the BG binary, $D = \sqrt{D^i D_i}$, and
\begingroup
\allowdisplaybreaks
\begin{subequations} \label{eq:simple terms}
    \begin{align}
        A_{1}^{ab} &= \ddddot{Q}_{\ub}^{ab} + \frac{2}{D}\dddot{Q}_{\ub}^{ab} + \frac{3}{D^2}\ddot{Q}_{\ub}^{ab} + \frac{3}{D^3}\dot{Q}_{\ub}^{ab} + \frac{3}{D^4} Q_{\ub}^{ab},\\
        A_{2}^{ab} &= 2\ddddot{Q}_{\ub}^{ab} + \frac{8}{D}\dddot{Q}_{\ub}^{ab} + \frac{18}{D^2}\ddot{Q}_{\ub}^{ab} + \frac{30}{D^3}\dot{Q}_{\ub}^{ab} + \frac{30}{D^4} Q_{\ub}^{ab},\\
        A_{3}^{ab} &= \ddddot{Q}_{\ub}^{ab} + \frac{10}{D}\dddot{Q}_{\ub}^{ab} + \frac{45}{D^2}\ddot{Q}_{\ub}^{ab} + \frac{105}{D^3}\dot{Q}_{\ub}^{ab} + \frac{105}{D^4} Q_{\ub}^{ab}.
    \end{align}
\end{subequations}
\endgroup

\paragraph{Toy Model---} For general $D^i$ the problem is analytically intractable. To illustrate resonant locking, we briefly present a toy model in the far-wave zone ($D \gg \lambdabar$, with $\lambdabar$ being the reduced wavelength of the BG binary's GWs), where the force reduces to \citep{Mashhoon_1978, Turner_1979, Rozneretal2020, Desjacquesetal2020}
\color{black}
\begin{equation}\label{eq:force on binary from GW}
    \frac{F^i}{\mu_\uf} = \ddot{r}^i_{\mathrm{f}} + \frac{M_{\mathrm{f}}}{r^3_{\mathrm{f}}} r^i_{\mathrm{f}} = \frac{1}{2} {\ddot{h}^{i}}_{j} r^j_{\mathrm{f}}, 
\end{equation}
where $h_{ij} = 2\ddot{Q}_{ij}/D$ is the GW strain of the BG binary. In the quadrupole approximation, $\psi = 2\phi$. For simplicity, we assume both binaries are circular.
\begin{table}[t]
        \def\sep{0.1ex}
        \begin{tabular}[c]{c|c}
            \multicolumn{2}{c}{} \\ [\sep]
            \hline\\[-2.3ex]
            $\phi$ & True anomaly.\\ [\sep]
            $n$ & Mean motion frequency.\\ [\sep]
            $v$ & Orbital velocity.\\ [\sep]
            $T$ & Orbital Period.\\ [\sep]
            $\Mchirp$ & Chirp mass.\\ [\sep]
            $M$ & Total mass.\\ [\sep]
            $\mu$ & Reduced mass.\\ [\sep]
            $\eta$ & Symmetric mass ratio.\\ [\sep]
            $r^i$ & Separation vector.\\ [\sep]
            $r$ & Separation vector length.\\ [\sep]
            $a$ & Semi-major axis.\\ [\sep]
            $e$ & Eccentricity.\\ [\sep]
            $\iota$ & Inclination.\\ [\sep]
            $\Omega$ & Longitude of the ascending node.\\ [\sep]
            $\omega$ & Argument of periapsis.\\ [\sep]
            $\varpi$ & Longitude of periapsis.\\ [\sep]
            $u$ & Mean anomaly.\\ [\sep]
            $\psi$ & Phase of the GW.\\ [\sep]
            $Q^{ij}$ & Quadrupole Moment.\\ [\sep]
            \hline
        \end{tabular}
        \caption{Definitions of binary parameters and variables used in this paper.}
    \label{tab:Parameters}
\end{table}


\paragraph{Evolution of the FG Binary---} The rate of work done due to GW stimulated emission or absorption is
\begin{align}
    \frac{\dd E}{\dd t} = W_{\mathrm{GW}} = \frac{\mu_{\mathrm{f}}}{2} \dot{r}_{\mathrm{f}}^i \ddot{h}_{ij} r_{\mathrm{f}}^j. \label{eq:Wse Defn}
\end{align}
In Cartesian coordinates the GW perturbation is \citep{Inverno_2022}
\begin{align}
    h_{ij} = 
    \begin{pmatrix}
        A_+ \cos(\psi_\ub) & A_\times \sin(\psi_\ub) & 0\\
        A_\times \sin(\psi_\ub) & - A_+ \cos(\psi_\ub) & 0\\
        0 & 0 & 0\\
    \end{pmatrix}, \label{eq:h_ij}
\end{align}
where $A_{+,\times}$ are the GW polarization amplitudes. For a circular BG source \citep{fesik_2017}
\begin{align}
    A_+ = -A_{\ub} (1 + \cos^2\chi), 
    \quad A_\times = -2 A_{\ub} \cos\chi, \label{eq:defn of GW polarization}
\end{align}
where $A_{\ub} = 2 \Mchirp_{\mathrm{b}}^{5/3} n_\ub^{2/3}/D$ and $\chi$ is the angle between the BG binary's orbital axis and $D^i$.

\black{We take $D^i$ to be  along the positive $\zhat$-axis, and the FG binary's orbit to be in an arbitrary plane}
\begin{align}
    r_\uf^i = \mathbf{O}_z(\omega_\uf)\mathbf{O}_x(\iota_\uf)
    \left(
        r_\uf \cos(\phi_\uf), \ 
        r_\uf \sin(\phi_\uf), \ 
        0 \right), \label{eq:defn of r_f}
\end{align}
where $\mathbf{O}_\alpha(\beta)$ are rotation matrices about the $\hat{\boldsymbol{\alpha}}$-axis by an angle $\beta$. Averaging Eq.~\eqref{eq:Wse Defn} over one FG orbit and keeping secular resonant terms yields
\begin{align}
    \frac{W_{\mathrm{GW}}}{4 \Mchirp_\uf^{5/3}  n_\uf^{-1/3} n_\ub^2 A_{\ub}} &\to \mathcal{A} \sin(2\phi_\uf - 2\phi_\ub + \vartheta), \label{eq:WGW simp}
\end{align}
where $\mathcal{A} (\iota_\uf, \omega_\uf, \chi)$ and $\vartheta (\iota_\uf, \omega_\uf, \chi)$ are given in Eqs.~\eqref{eq:mathcalA} and \eqref{eq:vartheta}. The energy of the FG binary system is, to leading order in the center-of-mass frame, 
\begin{align}
    E_\uf = \frac12 \mu_\uf \dot{r}_\uf^2 - \frac{M_\uf\mu_\uf}{r_\uf} = -\frac{\Mchirp^{5/3}_\uf n_\uf^{2/3}}{2}, \label{Vis-viva Eq.}
\end{align}
from Kepler's third law. Thus, $E_{\mathrm{f}} \propto n_{\mathrm{f}}^{2/3}$ and
\begin{align}
    \frac{\dot{n}_\uf}{n_\uf} = \frac{3}{2}\frac{\dot{E}_{\mathrm{f}}}{E_{\mathrm{f}}} = -\frac{3}{\Mchirp^{5/3}_\uf n_\uf^{2/3}} \dot{E}_{\mathrm{f}} \label{eq:log of energy and freq}.
\end{align}
Let us decompose $\dot{E}_\uf \equiv W_{\mathrm{GW}} + L_\uf$, where \citep{Peters_1963}
\begin{align}
        L_{\uf} = -\frac{32}{5} \frac{\Mchirp_{\uf}^5}{r_{\uf}^5 \eta_{\uf}} = -\frac{32}{5} (n_\uf \Mchirp_{\uf})^{10/3} \label{eq:grav lum of FG new}
\end{align}
is the FG binary's own GW loss. This gives
\begin{align}
    \begin{aligned}
        \dot{n}_\uf = \frac{96\Mchirp_{\mathrm{f}}^{5/3}}{5} n_\uf^{11/3} -24 \mathcal{A} \frac{\Mchirp_{\mathrm{b}}^{5/3}}{D} n_{\ub}^{8/3} \sin(2\phi_\uf - 2\phi_\ub + \vartheta), \label{eq:final 2nd Order ODE}
    \end{aligned}
\end{align}
where the BG binary's phase evolves as predicted by GR \citep{Maggiore_2008}
\begin{align}
    \phi_\ub &= \frac{n_{0,\ub}^{-5/3}}{32 \Mchirp_{\mathrm{b}}^{5/3}} \left[1 - \left(1 - \frac{256}{5} \Mchirp_{\mathrm{b}}^{5/3}n_{0,\ub}^{8/3} t\right)^{5/8}\right] + \phi_{0,\ub}, \label{eq:phi_b full}
\end{align}
to leading post-Newtonian order, with $\phi_{0,\ub}$ and $n_{0,\ub}$ being the initial phases and angular velocities, respectively. Eqs.~\eqref{eq:final 2nd Order ODE} and \eqref{eq:phi_b full} give a closed description of the FG binary's evolution over time. 


\paragraph{Forced Pendulum---} 
Defining the relative phase $\theta = 2\phi_\uf - 2\phi_\ub + \vartheta$\black{, neglecting GW emission from the FG binary (since $\Mchirp_\uf \ll \Mchirp_\ub$) and taking $\iota_\uf,\omega_\uf$, and $\chi$} fixed, Eq.\eqref{eq:final 2nd Order ODE} reduces to
\begin{align}
    \ddot{\theta} = - \CA \sin(\theta) - \CF, \label{eq:gen EOM}
\end{align}
where
\begingroup
\allowdisplaybreaks
\begin{subequations}\label{eq:Force coeff}
    \begin{align}
        \CA(t) &= 48 \mathcal{A}\frac{n_{\ub}^{8/3}(t) \Mchirp_{\ub}^{5/3}}{D} 
        = 24 \mathcal{A}\frac{\eta_\ub v_\ub^{5}(t)}{D \lambdabar(t)} > 0,\\
        \CF(t) &= \frac{192}{5} n_\ub^{11/3}(t) \Mchirp_{\mathrm{b}}^{5/3} = \frac{48 \eta_\ub  v_{\ub}^{5}(t)}{5 \lambdabar^2(t)} > 0.
    \end{align} 
\end{subequations}
\endgroup 
Eq.~\eqref{eq:gen EOM} is coincidentally the equation of motion of a forced pendulum with amplitude $\CA$ and forcing $\CF$.

A pendulum admits two types of motion: rotation ($2\pi$ cycles) and libration (oscillations about the stable equilibrium). Resonant locking corresponds to libration, where $\dot{\theta}$ remains bounded as $n_\ub$ and $n_\uf$ increase synchronously.

\paragraph{Resonant Locking \black{in the Toy Model}---} 

\begin{figure}[t]
    \centering
    \includegraphics[width=0.42\textwidth]{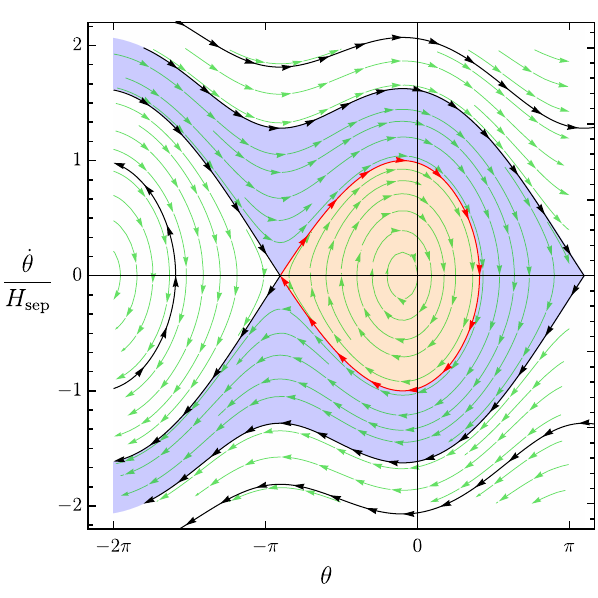}
     \caption{Phase-space of the forced pendulum (Eq.~\ref{eq:gen EOM}) with $\CF/\CA=0.3$. Closed librating contours are orange, open rotating contours are blue, the red separatrix separates these two regions and the outer separatrices are black. We have normalized $\dot{\theta}$ to the separatrix height, $H_{\rm sep}$.}
    \label{fig:Forced_pendulum_Separatrix}
\end{figure}

Locking requires a transition from rotation to libration. This process is governed by the phase-space structure of the forced pendulum (Figure~\ref{fig:Forced_pendulum_Separatrix}). From Eq.~\eqref{eq:Force coeff}, $\CA,\CF,$ and $\CF/\CA \propto n_\ub$ grow with time causing the separatrix to stretch vertically. Trajectories near it can be captured, moving the pendulum into libration and hence leading to locking. The initial phase $\theta_0$ dictates proximity to the separatrix at resonance, and a uniform $\theta_0 \in [0,2\pi)$ yields an (approximately) uniform distribution between the inner and outer separatrices at $\theta=-\pi$. This is illustrated in two videos: one showing capture, linked \href{https://youtu.be/xtwmT8TNJJA}{here}, and one showing no capture, linked \href{https://youtu.be/hWVE7h16_AU}{here}.

Suppose the system enters libration at frequency $n_{\mathrm{b,lock}}$, and reduced wavelength $\lambdabar_{\mathrm{lock}} = 1/(2n_{\mathrm{b,lock}})$, in the wave zone. The FG binary will remain in the wave zone provided $D$ is fixed, and $n_\uf$ will oscillate about $n_\ub$ provided $\CF < \CA$. Since $\CF/\CA \propto n_\ub$, libration will inevitably end when $\CF > \CA$, after which $\ddot{\theta}<0$ permanently and the system resumes rotation. Thus, during locking,
\begin{align} 
    \frac{\CF}{\CA} = \frac{4}{5 \mathcal{A}} D n_\ub = \frac{2 D_{\lambdabar_{\mathrm{lock}}}}{5 \mathcal{A}} \frac{n_\ub}{n_{\mathrm{lock}}} \leq 1, \label{eq:C_F < C_A} 
\end{align}
where $D_{\lambdabar} = D/\lambdabar$. Since $\max(\mathcal{A})=1$ for $(\iota_\uf,\chi) \in [0,\pi)$, we require $D_\lambdabar < 5/2$. This contradicts the $D \gg \lambdabar$ limit, confirming that a full EoM treatment in the $D \sim \lambdabar$ regime is necessary. Nonetheless, this toy model is valuable for qualitatively illustrating resonant locking.

\paragraph{Implication: producing merging systems---}
Resonant locking increases $n_\uf$, reducing the FG binary's merger time. For the sake of the argument, consider $\mathcal{A} = 1$. The FG binary's time to coalescence is \citep{Maggiore_2008}
\begin{align}
    t_{\rm merge, \uf} (n_{\uf})= \frac{5}{256} \Mchirp_\uf^{-5/3}n_{\uf}^{-8/3}. \label{eq:inspiral time}
\end{align}
If locking starts at $n_{\mathrm{start}}$ and terminates at frequency $n_{\mathrm{end}}$ (where $\CF = \CA$), then
\begin{align}
    \frac{t_{\rm merge,\uf}(n_{\mathrm{start}})}{t_{\rm merge,\uf}(n_{\mathrm{end}})} = \left(\frac{n_{\mathrm{end}}}{n_{\mathrm{start}}}\right)^{8/3} = \left(\frac{5}{2D_{\lambdabar}}\right)^{8/3}. \label{eq:reduced merg time}
\end{align}
\black{To illustrate how resonant locking can drive binaries towards merger, consider a system with $M_\uf = 40 \Msun$, $r_\uf = 0.5$ AU, and $D_{\lambdabar} = 1$, which gives $D = 100$ AU. During locking, the orbital period goes from 7.24 days to 2.90 days, and the merger time is brought down from $5.76 \ T_{\rm H}$ to $0.50 \ T_{\rm H}$, where $T_{\rm H}$ is a Hubble time.} Thus, binaries that would not merge within $T_{\rm H}$ via GW emission alone can do so after resonant locking with a faster-inspiralling BG binary. This may have important implications for GW population-synthesis models.

\color{black} 

We now turn to the full treatment using the general force in Eq.~\eqref{eq:Full force}.

\paragraph{General EoM---} 
When $D \sim \lambdabar$, each term in Eq.~\eqref{eq:simple terms} contributes approximately equally, so the general expression is required. We also include the FG binary’s orbit about the BG binary (the `outer' orbit), giving a total of three intrinsic Euler angle triplets $(\Omega_i, \iota_i, \omega_i)$ with $i \in \{\rm f,b, outer\}$. Orbital elements are allowed to vary self-consistently via the Lagrange planetary equations (LPEs; Eq.~\eqref{eq:LPEs wrt M} in Appendix and surrounding text) under the influence of the full perturbing force -- Eq.~\eqref{eq:Full force}. We focus on low-eccentricity systems ($e \lesssim 0.1$), leaving the general eccentric case for future work.

This general treatment raises two questions: (1) Is resonant locking still possible? (2) If so, how does capture occur?

\begin{figure}[t]
    \centering
    \includegraphics[width=0.49\textwidth]{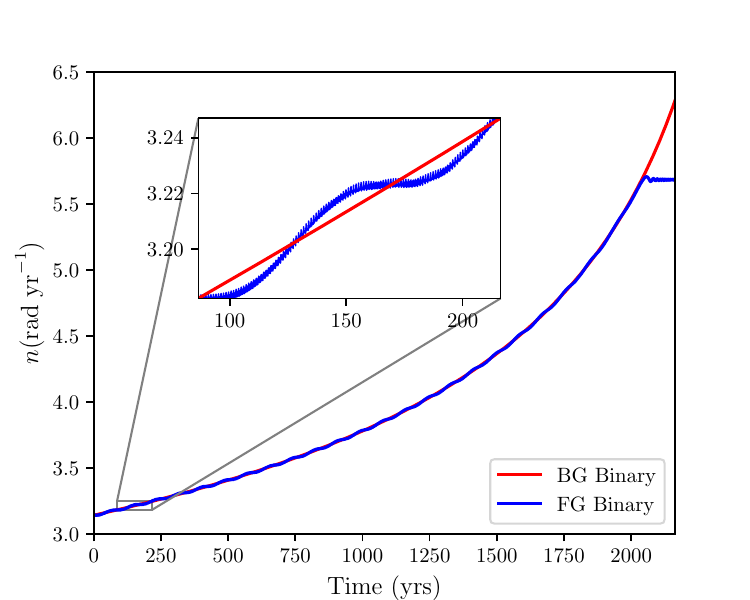}
     \caption{
     \color{black} 
     We numerically integrate Eq.~\eqref{eq:LPEs wrt M} with the general force (Eq.~\ref{eq:Full force}) using initial conditions $\mathcal{M}_\ub = 8.71 \times 10^8 M_{\odot}, \Dlb = 1, e_{\ub,0} = 0, \phi_{\rm b,0} = 0, T_{\rm b,0} = T_{\rm f,0} = 2 {\rm \ yrs}, \phi_{\rm f,0} = 0.65 \pi, e_{\uf,0} = 0.1, \iota_\uf = 0, \iota_{\rm outer} = 0$ and all other orbital elements are initially set to zero. This gives $T_{\rm outer} \approx 22.6$ yr. The FG binary’s mean motion $n_\uf$ remains trapped in resonance until decoupling at $T_\uf \approx 1.11$ yr.
     }
    \label{fig:example_locking}
\end{figure}

Figure~\ref{fig:example_locking} demonstrates locking is indeed possible for a non-circular FG binary (initially $e_{\rm f} = 0.1$ and remains approximately constant). Starting at resonance, $n_\uf$ oscillates slowly about $n_\ub$ due to secular terms, while faster oscillations arise from non-secular terms. 

In fact, we have found that it is possible for locking to begin in the near-zone and persist as the FG binary moves into the wave-zone. This is shown in Figure~\ref{fig:example_locking_D=0.5} and is discussed further in the Appendix. However, since our primary interest lies in GW-induced resonances, we confine our study to the wave zone and defer an analysis of near-zone resonances to future work.

We discuss below the conditions needed for generic FG/BG/outer binary configurations to produce resonant locking like that shown in Figure~\ref{fig:example_locking}.

During locking, the GW must increase the FG binary's mean-motion frequency at the same rate that the BG binary's frequency changes, i.e. $\dot{n}_\uf \geq \dot{n}_\ub$. If this does not occur, the GW will outrun the binary and resonance cannot be sustained. Since the GW interaction strength varies along the outer orbit, and the BG binary's inspiral timescale is much longer than $T_{\rm outer}$, we must outer-orbit-average. We arrive at the condition
\begin{align}
    \overline{\dot{n}}_\uf &=  -\frac{3n_\uf}{2a_\uf} \overline{\dot{a}}_\uf = -\frac{3n_\uf}{2a_\uf} \frac{1}{T_{\rm outer}}\int_0^{T_{\rm outer}} \dot{a}_\uf \ \dd t \ \geq \dot{n}_\ub, \label{eq:average change in FG frequency}
\end{align}
where $\dot{a}_\uf$ comes from Eq.~\eqref{eq:LPE for a}. Since the relative phase between the FG binary and BG binary controls whether energy is given or taken from the FG binary, we maximize $\overline{\dot{n}}_\uf$ over the initial relative phase between the FG and BG binaries. In other words, if, at resonance, there exists an initial relative phase that satisfies Eq.~\eqref{eq:average change in FG frequency}, then locking is possible.

Note that while $\dot{n}_\uf$ can fall below $\dot{n}_\ub$ during parts of the outer orbit, the slow drift of the GW frequency ensures the system remains within the resonance bandwidth over one outer orbital period and so it is the averaged rate of change that governs locking. This condition is both sufficient and necessary for resonant locking, provided the binary has a suitable initial phase relative to the GW. To verify this numerically, we randomly sampled system parameters and found that all cases satisfying $\overline{\dot{n}}_\uf \geq \dot{n}_\ub$ indeed achieved resonant locking, including extremal values where $\overline{\dot{n}}_\uf/\dot{n}_\ub$ is close to unity and $\overline{\dot{n}}_\uf/\dot{n}_\ub$ is much larger than unity.

Our system has 12 initial degrees of freedom (9 Euler angles and 3 phases). To identify which are important for low-eccentricity orbits, it is instructive to consider first the circular case ($e=0$), since our results remain essentially unchanged for small $e$\footnote{\black{We confirm using numerical simulations as in Figure~\ref{fig:example_locking}.}}. In this limit, 
only three non-degenerate parameters affect the evolution: $\iota_\uf$ and $\iota_{\rm outer}$ relative to the BG binary's orbital plane, and the relative phase $\Delta\phi_0 = \phi_{{\rm f},0} - \phi_{{\rm b},0}$. Note that we must extend the domain of inclinations from $[0,\pi)$ to $[0,2\pi)$ in order to absorb the degeneracies with longitudes $\Omega_{\rm outer}$ and $\Omega_{\rm f}$. The inclinations also obey a $(\iota_\uf, \iota_{\rm outer}) \to (2\pi - \iota_\uf,2\pi - \iota_{\rm outer})$ symmetry.

Resonant locking is possible if, for given $\iota_\uf$ and $\iota_{\rm outer}$, there exists a $\Delta\phi_0$ such that $\overline{\dot{n}}_\uf \geq \dot{n}_\ub$. This will of course depend on $\Dlb$.

\begin{figure}[t]
    \centering
    \includegraphics[width=0.49\textwidth]{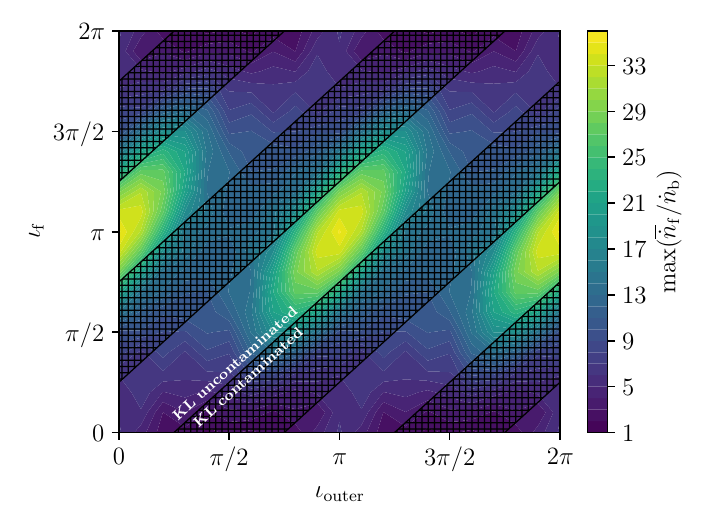}
     \caption{
     \color{black} 
     The maximum value of $\overline{\dot{n_\uf}}/\dot{n}_\ub$ when varying the relative phase of the FG and BG binaries as a function of the inclinations $\iota_\uf$ and $\iota_{\rm outer}$ relative to the BG binary for $D = \lambdabar$. The hatched regions represent the parameter space removed due to ZLK oscillations. The condition for resonance capture, $\overline{\dot{n_\uf}}/\dot{n}_\ub > 1$, is satisfied for all $\iota_\uf$ and $\iota_{\rm outer}$.
     \color{black}
     }
    \label{fig:f_max_KL_contamination_T12345_Dlb=1}
\end{figure}

In the toy model discussed above, only a single resonance appeared, at $n_\uf = n_\ub$. Now that we allow the FG binary to orbit the BG binary, we get a total of five distinct resonances at $n_\uf = n_\ub + k n_{\rm outer}$ for $k \in \{-2,-1,0,1,2\}$%
\footnote{\black{Allowing the FG binary to orbit the BG binary introduces new terms proportional to $\cos(2(\phi_\uf - \phi_\ub + k \phi_{\rm outer}))$, where $k \in \{-2,-1,0,1,2\}$, when expanding Eq.~\eqref{eq:Full force}.} }. These resonances do not overlap and their strengths depend on the Euler angles. 
Figure~\ref{fig:f_max_KL_contamination_T12345_Dlb=1} shows the maximum of $\overline{\dot{n_\uf}}/ \dot{n}_\ub$ across all five resonances, when $\Delta \phi_0$ is varied, as a function of $\iota_{\rm outer}$ and $\iota_\uf$ for $D_{\lambdabar} = 1$. We find values between 1.7 and 35, implying that locking can occur for any orbital configuration at this distance. It is around $\Dlb \gtrsim 2$, however, that $\overline{\dot{n_\uf}}/ \dot{n}_\ub < 1$ everywhere (the exact distance depends on $\iota_\uf$ and $\iota_{\rm outer}$ -- see Appendix Figure~\ref{fig:f_max_KL_contamination_T12345_Dlb=2}).

The remaining challenge is resonance capture: acquiring the phase needed for locking. In our toy circular-restricted model (Eq.~\ref{eq:gen EOM}), a FG binary with a suitable initial relative phase could lock with the GWs by crossing the separatrix. In the general treatment, however, variations in the FG binary’s longitude of periapsis, $\varpi=\omega+\Omega$, inhibit separatrix crossing and thus resonance capture.

We identify two mechanisms that can lead to locking, both confirmed numerically. One is separatrix crossing with the additional requirement that the outer orbit's separation distance, $a_{\rm outer}$, must be decreasing. Specifically we require $\dot{a}_{\rm outer}/a_{\rm outer} \gtrsim \dot{n}_\ub/n_\ub$%
\footnote{\black{This was verified numerically, and is identical to the condition that the separatrix increases in both height \textit{and} width in the $D \gg \lambdabar$ regime.}}. Once captured, locking can persist even if $a_{\rm outer}$ stops decreasing. A dynamical process that could cause such behaviour of $a_{\rm outer}$ is the ``tidal'' disruption of a wide triple near the periapsis of its eccentric orbit around the BG, thus depositing the inner binary of the triple, the FG binary, on a more tightly bound orbit about the BG binary than it originally was. We leave this mechanism for future work. 

The second mechanism is getting kicked into resonance via dynamical scattering encounters -- the primary process we consider next. The nature of these encounters of course depends on the binary's astrophysical environment.

\paragraph{Post-Merger Galaxies---}
Since capture is probabilistic, resonant locking is most effective in dense environments containing many stellar-mass binary objects (smBOs) around a strong GW source. A natural setting is in the bulge of a post-galaxy merger, where the SMBHB forms our BG binary and the surrounding smBOs act as FG binaries (see Appendix for details of the galaxy/bulge model). As the SMBHB inspirals, it sweeps through almost all relevant frequencies, giving each smBO an opportunity to be captured. We model the kicks from scattering encounters and estimate the resulting number of resonantly locked systems.

\paragraph{Kicked into Resonance by Dynamical Encounters---} Dynamical encounters, which represent the dominant noise source in galactic bulges, can knock the FG binary either \textit{into} resonance or knock it \textit{out} of resonance. Although the aforementioned separatrix approach is not strictly valid in our full treatment, we will apply it here as it provides a transparent and intuitive framework for gaining an order-of-magnitude estimate of this stochastic capture rate. We note that the resonances in this general treatment are of similar strength to, and are sometimes stronger than, the resonance in the toy model and hence our estimate is a conservative underestimate.

Consider an encounter between the FG binary and a perturber of mass $M_{\mathrm{p}}$, impact parameter $b$, and velocity $v$, taken as the bulge velocity dispersion from the $M_{\rm SMBH}$–$\sigma$ relation \cite[][Eq.~(7)]{Kormendy_2013}:
\begin{align}
    v = \sigma = 200 {\rm km/s} \left(\frac{M_{\rm SMBH}}{3.08 \times 10^8 M_{\odot}}\right)^{1/4.38}. \label{M-sigma relation}
\end{align}
There are two regimes to consider: slow encounters, where $(v/b) T_\mathrm{f} < \gamma$, and fast encounters, where $(v/b) T_\mathrm{f} > \gamma$, with $\gamma = \mathcal{O}(1)$ (the exact value depends on one's definition of `fast' and `slow' and the results are insensitive to it). During a slow encounter, the binary rotates rapidly, and the work done by the perturbing star averages to zero. Such encounters hence cannot push the binary into/out of resonance. In the fast regime the phase is effectively frozen during the interaction and the perturber delivers a net kick, changing the FG binary's energy by \cite[][Eq.~(8.42)]{binney_2008}
\begin{align}
    \Delta E \sim \frac{G^2 M_{\mathrm{p}}^2 M_\uf r_\uf^2}{2v^2 b^4} = \frac{G^{8/3} M_{\mathrm{p}}^2 M_\uf^{5/3}}{2 n_\uf^{4/3} v^2 b^4}. \label{eq:Delta E from encounter}
\end{align}
where $G$ is the gravitational constant. This can be negative or positive depending on the phase. These fast encounters are the ones relevant for kicks into resonance.

Further suppose that the FG binary is not in resonance during a fast encounter and that $\Gamma_{\rm into}(T_\uf,D,t)$ is the rate of kicks that are just the right strength to knock the binary system inside the separatrix. The probability density that such a kick first occurs at time $t$ follows the exponential density distribution of a Poisson process
\begin{align}
    \mathcal{P}_{\rm into} (T_\uf, D, t_1, t) = \Gamma_{\rm into} \exp \left[- \int_{t_1}^{t} \Gamma_{\rm into} \dd t'\right],
\end{align}
where $t_1(M_\ub)$ marks the time the FG binary first enters the wave-zone; $\lambdabar(t_1) = D$. The full expression for $\Gamma_{\rm into}$ is given in Eq.~\eqref{eq:gamma_into expression}.

Resonance must then persist until natural decoupling, i.e.~until $\overline{\dot{n}_{\uf}}/\dot{n}_\ub$ drops below unity. We take this to be at $D = 2\lambdabar$ (the exact distance depends on $\iota_\uf$ and $\iota_{\rm outer}$ and typically lies between $1.5\lambdabar$ and $2.3\lambdabar$). If $\Gamma_{\rm out}(t, D)$ is the rate of kicks large enough to remove the binary from resonance (i.e.~a kick with $2\Delta n_\uf > H_{\rm sep}$), the survival probability of a system that was captured at time $t$ is
\begin{align}
    P_{\rm stable} &= \exp \left[{-\int_{t}^{t_2}\Gamma_{\rm out} \dd t'} \right], \label{eq:P_stable}
\end{align}
where $t_2$ marks the time the FG binary naturally decouples, defined implicitly through $\lambdabar(t_2) = D/2$. The full expression for $\Gamma_{\rm out}$ is given in Eq.~\eqref{eq:gamma_out}.

Other dynamical processes, such as the von Zeipel–Lidov–Kozai (ZLK) effect \cite{Lidov_1962,Kozai_1962}, can interfere with resonant locking and we must thus constrain our parameter space. In particular, ZLK is important for $40^{\circ} \lesssim \iota_\uf - \iota_{\rm outer} \ ({\rm mod \ }180^{\circ}) \lesssim 140^{\circ}$ \citep{Naoz_2016} as it induces eccentricity oscillations. We therefore exclude this inclination range. Note, however, that ZLK oscillations do not change the FG binary's orbital frequency and will thus not necessarily disrupt locking even if oscillations are large -- however, we leave this to future work. Assuming inclinations are uniform in $\cos i$, this removes 77.5\% of parameter space, so we multiply $P_{\rm locked}$ by 0.225.

Further, recall that in the general treatment, we found non-overlapping resonances at $n_\uf = n_\ub + k n_{\rm outer}$ for $k \in \{-2,-1,0,1,2\}$. We show in the Appendix, namely in Figure~\ref{fig:nf/nb for general treatment}, that all 5 resonances are of similar strength. Consequently, locking can occur in identical fashion at each resonance. For concreteness, we also demonstrate that resonant locking is possible at each resonance in Figure~\ref{fig:example_locking_diff_res}. Some (in fact, most) regions of $(\iota_\uf, \iota_{\rm outer})$ space are home to multiple of these resonances (cf. Figure~\ref{subfig:num_reses}). On average, each configuration with $\iota_\uf - \iota_{\rm outer} \ ({\rm mod \ }180^{\circ}) \lesssim 40^{\circ}$, or $\iota_\uf - \iota_{\rm outer} \ ({\rm mod \ }180^{\circ}) \gtrsim 140^{\circ}$ hosts 2.59 resonances%
\footnote{\black{This is calculated by averaging the number of resonances shown in Figure~\ref{subfig:num_reses} across the non-ZLK contaminated parameter space.}}. We hence multiply $P_{\rm locked}$ by 2.59.

The probability of being locked is therefore
\begin{align}
    P_{\rm locked} (T_\uf, D) &= 2.59 \times 0.225 \int_{t_1}^{t_2} \mathcal{P}_{\rm into} P_{\rm stable} \ \dd t,
\end{align}
which depends on the FG binary’s initial period and distance from the SMBHB, the BG binary’s mass, and the perturber mass, which we take to be the average star mass $\langle m_{\mathrm{star}} \rangle = 0.3 M_{\odot}$ \citep{Kroupa2002}.

\paragraph{The Number of GW Locked Systems---}
The cumulative number of binaries that lock and remain locked until natural decoupling during a galaxy merger is
\begin{align}
    &\frac{N_{\mathrm{locked}}}{N_{\mathrm{bin}}} = \int_{0}^{\infty} \int_{T_{\mathrm{min}}}^{T_{\mathrm{max}}} P_{\rm locked}(D,\lambdabar) f_{\mathrm{D}}(D) f_{T}(T) \dd T \dd D , \label{eq:num of locked binaries}
\end{align}
where $\lambdabar(T) = cT/(4\pi)$, $f_D(D)$ is the radial distribution, $f_T(T)$ is the period distribution, and $N_{\rm bin}=0.25N_{\rm stars}$ is the number of binaries in the bulge \cite{Gao_2014, Boffin_2025} where $N_{\rm stars} = M_{\mathrm{bulge}}/\langle m_{\mathrm{star}} \rangle$ and $M_{\mathrm{bulge}}$ comes from the $M_{\rm bulge}$--$M_{\rm SMBH}$ relation (Eq.~\ref{eq:Mbulge MSMBH rel}). See Appendix for details, including the integration bounds.

\begin{figure}[t]
    \centering
    \includegraphics[width=0.42\textwidth]{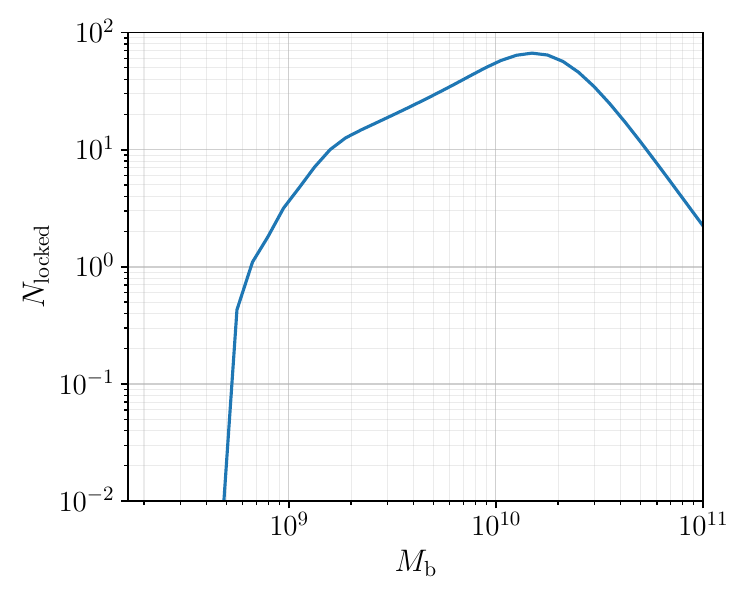}
     \caption{\black{The number of locked binary systems as a function of $M_\ub$, using Eq.~\eqref{eq:num of locked binaries}.}}
    \label{fig:number_locked}
\end{figure}

Figure~\ref{fig:number_locked} shows $N_{\rm locked}$ for varying $M_\ub$. It peaks at $M_\ub \sim 10^{10} M_\odot$ due to competing effects. At small $M_\ub$, higher central densities shorten evaporation times (Eq.~\ref{eq:evap time}), reducing $T_{\max}$ and hence bottlenecking $N_{\rm locked}$ for $M_\ub \lesssim 10^{8.5} M_\odot$. At larger $M_\ub$, the BG binary's inspiral accelerates, so FG binaries spend less time in the $1 < \Dlb < 2$ zone, increasing $P_{\rm stable}$. At very large $M_\ub > 10^{10} M_\odot$, the inspiral is too rapid for binaries to be knocked into resonance in the first place, lowering $\mathcal{P}_{\rm into}$. The result is a peak near $10^{10} M\odot$ where these effects balance.

\paragraph{Discussion---} We find that up to $\mathcal{O}(30)$ binaries can become resonantly locked with a SMBH binary during a high-mass galaxy merger (Figure~\ref{fig:number_locked}). Further, using the $M_{\rm SMBH}$–$M_{\rm DM}$ relation \citep[][Eq.~(18)]{Kormendy_2013} and integrating over the DM halo mass function \citep[][Eq.~(10)]{Kormendy_2013}, we estimate that $\mathcal{O}(10^{10})$ binaries have been locked at some point during the cosmic history of the observable Universe (see Appendix).

A direct signature of locking could be anomalous inspiral detected either by EM observations of binary pulsars or GW observations of compact objects. Future detectors could observe a resonantly locked binary inspiraling more quickly than what is predicted by GR. This would be a smoking-gun signature of locking.

There are limitations to the work discussed here. We have only considered perturbations to the binary from passing stars when there could be other local effects (such as the presence of planets/gas) that influence the dynamics. We have not studied binaries with large eccentricities. However, given that an eccentric orbit may be decomposed in Fourier modes \cite{Hui_2013}, we would expect similar behaviour except with additional harmonics. The combined effect of a resonant GW interaction in the presence of ZLK oscillations with inclinations $40^{\circ} < \iota_\uf - \iota_{\rm outer}\ ({\rm mod \ }180^{\circ}) < 140^{\circ}$ has been ignored. We ignore near-zone resonant locking. We wish to explore these directions in future work, as well as the possibility of entering the resonance via a ``tidal'' disruption of a hierarchical triple, on an initially eccentric outer orbit, as alluded to above.

\paragraph{Conclusion---} We have studied a previously unrecognized dynamical mechanism: the resonant locking of stellar-mass binaries with gravitational waves generated by supermassive black hole mergers. We expect this phenomenon to take place in high mass post-merger galaxies, with billions of BH binaries being resonantly locked in the Universe's history.

\color{black}

\begin{acknowledgments}
    We are grateful to Avi Loeb for related initial discussions which motivated this work, to John Magorrian for helpful discussions on forced pendulums\black{, to Adrien Kuntz for discussions about the near-zone effects, and to Thomas Spieksma for sharing his knowledge about scalar boson clouds affecting binary inspirals}. This work was supported by the UK Science and Technology Facilities Council Grant Number ST/W000903/1 \black{and} by a Leverhulme Trust International Professorship Grant (No.~LIP-2020-014). Y.B.G.~was partly supported by the Simons Foundation via a Simons Investigator Award to A.A.~Schekochihin. Z.X.~acknowledges partial support from the Bhaumik Institute for Theoretical Physics summer fellowship, NSF-AST Grant No.~2206428, and thanks Howard and Astrid Preston for their generous support.
\end{acknowledgments}

\bibliographystyle{apsrev4-2}
\bibliography{apssamp}

\section{Appendix}

\begin{figure}[ht]
    \centering
    \begin{tikzpicture}[scale=1.2]
        \def\R{2.5}     
        \def\BGWZr{1.8} 
        \def\FGorb{0.25}   
        \def\BGorb{0.4}   
        \def\BGsize{3.5pt}
        \def\FGsize{1.5pt}
        
        \coordinate (FG) at (0,0);    
        \coordinate (BG) at (0,-\R);    
        
        \draw (FG) ellipse({\FGorb} and {0.25*\FGorb});
        \fill ($(FG) + (-1*\FGorb,0)$) circle(\FGsize);
        \fill ($(FG) + (1*\FGorb,0)$) circle(\FGsize);
        
        \draw[dotted,thick] (BG) circle(\BGWZr);
        \draw (BG) ellipse({\BGorb} and {0.25*\BGorb});
        \fill ($(BG) + (-1*\BGorb,0)$) circle(\BGsize);
        \fill ($(BG) + (1*\BGorb,0)$) circle(\BGsize);
        
        \draw[<->] ($(BG) + (0,0.13*\BGWZr)$) -- ($(BG) + (-0.95*\BGWZr,0.13*\BGWZr)$) 
        node[midway,above]{$\lambdabar$};
        \draw[<->] ($(BG) + (2*\BGorb,0)$) -- ($(2*\BGorb,0)$) 
        node[midway,right]{$D$};
        
        \draw[decorate,decoration={snake,amplitude=1.2mm,segment length=10mm},thick,-{Stealth[length=3.5mm,width=2.5mm,sep=-1pt]}] ($(BG) + (0,0.6)$) -- ($(FG) + (0,-0.3)$)
        node[midway,left,xshift=-3pt]{GW};

        \node at ($(BG) + (-0.4,-0.4)$) {BG};
        \node at ($(FG) + (-0.4,0.3)$) {FG};
        
    \end{tikzpicture}
    \caption{\black{Illustration of the setup. The dotted line represents the wave-zone/near-zone boundary. The BG binary is emitting GWs in all directions. Note that, despite the depiction, these GWs actually form near the wave-zone/near-zone boundary rather than inside the near zone.}}
    \label{fig:systematic_setup}
\end{figure}

\color{black} 

\paragraph{Expressions for $\mathcal{A}$ and $\vartheta$---}
The expressions for $\mathcal{A}$ and $\vartheta$ in Eq.~\eqref{eq:WGW simp} are
 \begin{align}
     \mathcal{A} &= \frac{1}{8}\sqrt{4(C_{\iota_\uf} + C_{\chi})^2(1 + C_{\iota_\uf} C_{\chi})^2 + S_{\chi}^4 S_{\iota_\uf}^4\cos(2\omega_\uf)}, \label{eq:mathcalA}\\
     \vartheta &= \pi + \arctan \left(\frac{1 - t_{\chi}^4 t_{\iota_\uf}^{4}}{1 + t_{\chi}^4 t_{\iota_\uf}^{4}} \tan(2\omega_\uf)\right), \label{eq:vartheta}
 \end{align} 
respectively, with $t_{x} = \tan(x/2)$, $S_{x} = \sin(x)$ and $C_{x} = \cos(x)$ for $x \in \{\chi, \iota_\uf\}$.

\begin{figure}[t]
    \centering
    \includegraphics[width=0.49\textwidth]{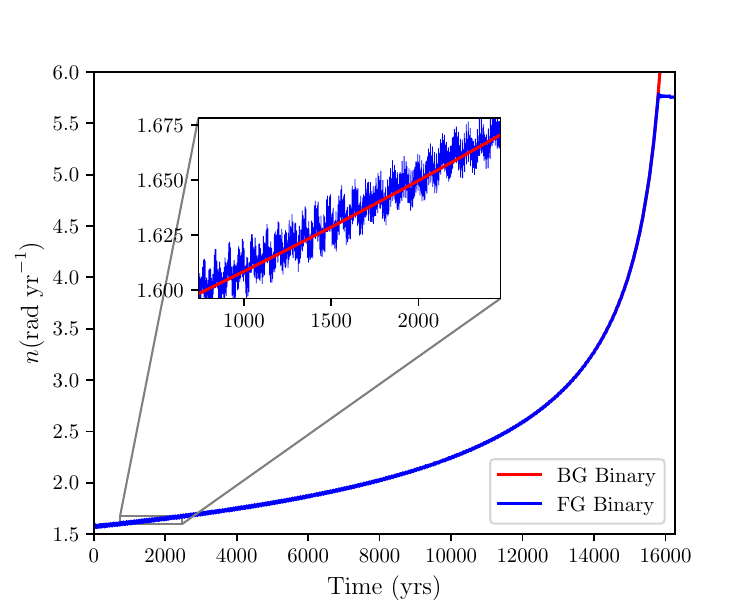}
    \caption{
     \color{black} 
     We demonstrate sustained resonant locking as the FG binary crosses the near-zone/wave-zone boundary. Specificall, we numerically integrate Eq.~\eqref{eq:LPEs wrt M} with the general force (Eq.~\ref{eq:Full force}) using initial conditions $\mathcal{M}_\ub = 8.71 \times 10^8 M_{\odot}, \Dlb = 0.5, e_{\ub,0} = 0, \phi_{\rm b,0} = 0, T_{\rm b,0} = T_{\rm f,0} = 4 {\rm \ yrs}, \phi_{\rm f,0} = 0.65 \pi, e_{\uf,0} = 0.1, \iota_\uf = 0, \iota_{\rm outer} = 0$ and all other orbital elements are initially set to zero. This gives $T_{\rm outer} \approx 22.6$ yr. The FG binary’s mean motion $n_\uf$ remains trapped in resonance until decoupling at $T_\uf \approx 1.09$ yr.
     }
    \label{fig:example_locking_D=0.5}
\end{figure}

\begin{figure}[t]
    \centering
    \includegraphics[width=0.49\textwidth]{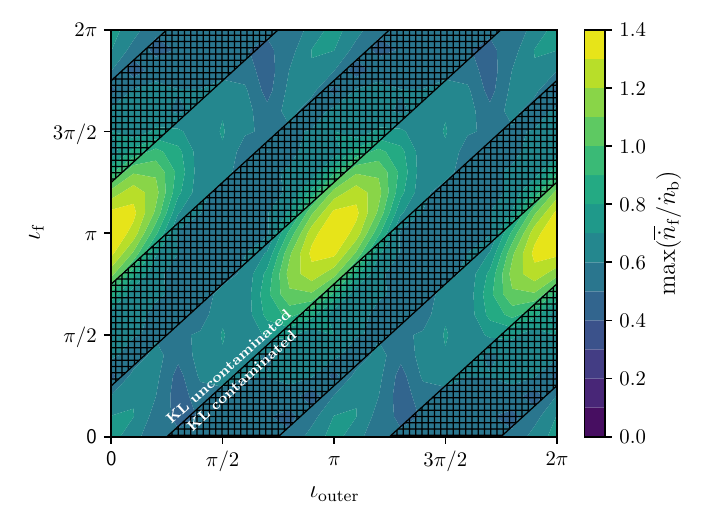}
     \caption{
     \color{black} 
     The maximum value of $\overline{\dot{n_\uf}}/\dot{n}_\ub$ due to the BG binary when marginalizing over all relative phases as a function of the FG orbit and outer orbit inclinations for $D = 2 \lambdabar$. The hatched regions represent the parameter space removed due to ZLK oscillations. At this distance, $\overline{\dot{n_\uf}}/\dot{n}_\ub$ does not exceed 1.4.
     \color{black}
     }
    \label{fig:f_max_KL_contamination_T12345_Dlb=2}
\end{figure}

\begin{figure*}[hbt!]
    \centering
        \subcaptionbox{$n_\uf = n_\ub + 2n_{\rm outer}$.
        \label{subfig:f_+2outer_0FGT12345}}
        {\includegraphics[width=.33\linewidth]{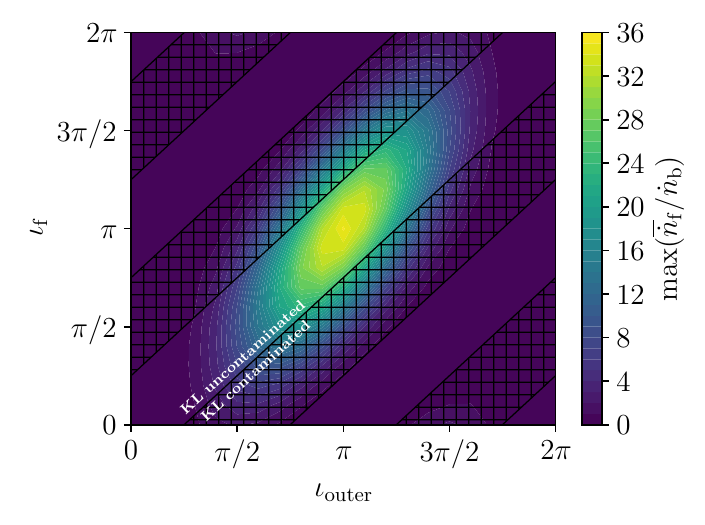}}\hfill 
        \subcaptionbox{$n_\uf = n_\ub + n_{\rm outer}$.
        \label{subfig:f_+1outer_0FGT12345}}
        {\includegraphics[width=.33\linewidth]{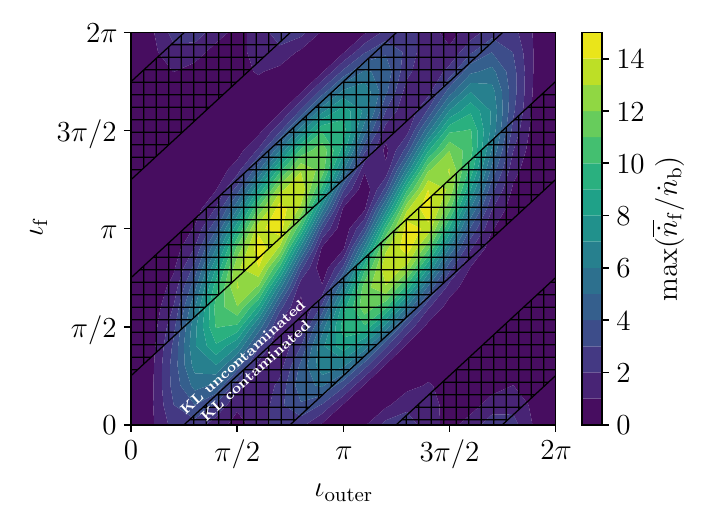}}\hfill 
        \subcaptionbox{$n_\uf = n_\ub$.
        \label{subfig:f_+0outer_0FGT12345}}
        {\includegraphics[width=.33\linewidth]{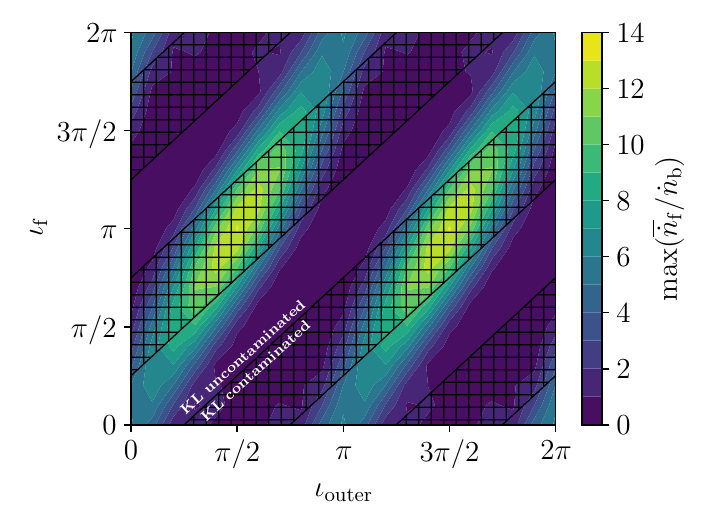}}\\
        \subcaptionbox{$n_\uf = n_\ub - n_{\rm outer}$.
        \label{subfig:f_-1outer_0FGT12345}}
        {\includegraphics[width=.33\linewidth]{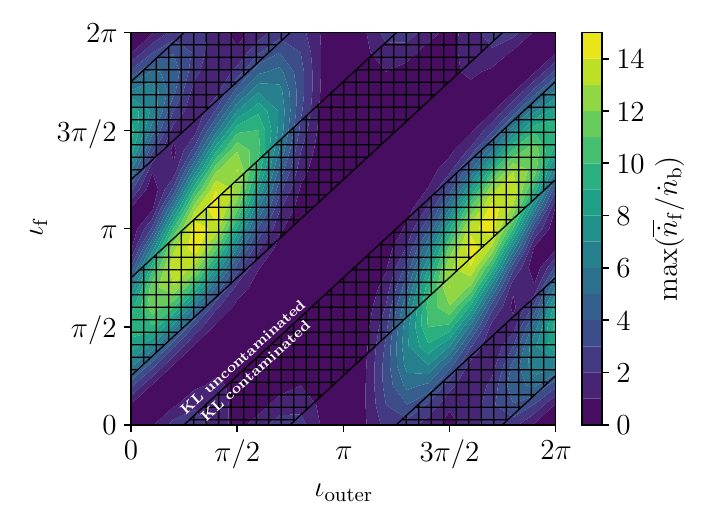}}\hfill 
        \subcaptionbox{$n_\uf = n_\ub - 2n_{\rm outer}$.
        \label{subfig:f_-2outer_0FGT12345}}
        {\includegraphics[width=.33\linewidth]{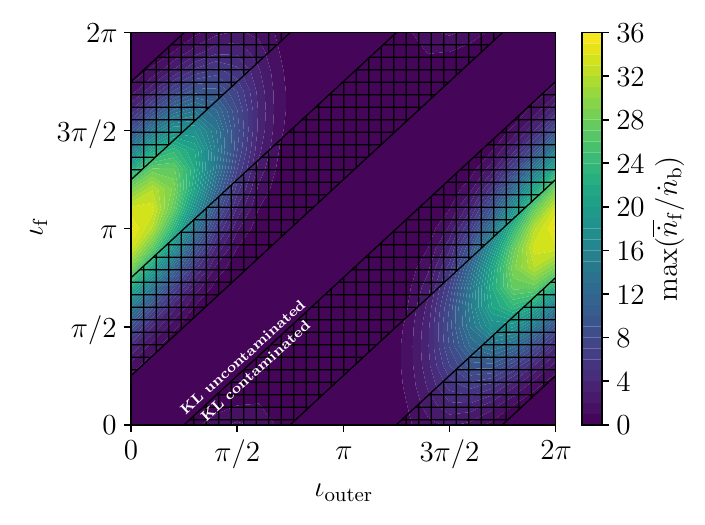}}\hfill 
        \subcaptionbox{The number of resonances for which $\max(\overline{\dot{n}}_\uf/\dot{n}_\ub) > 1$.
        \label{subfig:num_reses}}
        {\includegraphics[width=.33\linewidth]{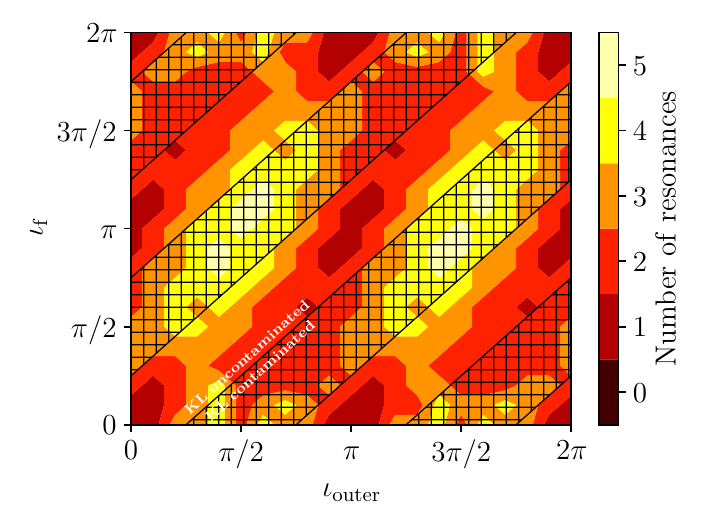}}\hfill 
    \caption{\black{The maximum value of $\overline{\dot{n}}_\uf/\dot{n}_\ub$ are shown in panels (a)-(e) for all 5 resonances, while panel (f) shows the total number of resonances where $\max(\overline{\dot{n}}_\uf/\dot{n}_\ub) > 1$. The hatched regions represent the parameter space removed due to ZLK oscillations.}}
    \label{fig:nf/nb for general treatment}
\end{figure*}

\paragraph{The Lagrange Planetary Equations---}
The LPEs are a set of 6 coupled first-order differential equations that describe how the 6 orbital elements ($a$, $e$, $\iota$, $u$, $\omega$, and $\Omega$) of a binary system evolve in the presence of an external perturbation \citep[][\S1.1]{Brumberg_1991}\citep[][\S11.9]{Danby_1962}\citep[][\S12.2]{Bertotti_2012}
\begingroup
\allowdisplaybreaks
\begin{subequations} \label{eq:LPEs wrt M}
    \begin{align}
        \dot{a} & = \sqrt{\frac{4 a}{G M}} \frac{\partial \mathcal{L}_{\rm p}}{\partial M}, \label{eq:LPE for a}\\
        \dot{e} & = -\sqrt{\frac{1-e^2}{G M a e^2}} \frac{\partial \mathcal{L}_{\rm p}}{\partial \omega}+\frac{1-e^2}{\sqrt{G M a e^2}} \frac{\partial \tilde{\mathcal{L}}_{\rm p}}{\partial M}, \\
        \frac{\mathrm{d}\iota}{\mathrm{d}t} & =-\frac{\csc \iota}{\sqrt{G M a\left(1-e^2\right)}} \frac{\partial \mathcal{L}_{\rm p}}{\partial \Omega} + \frac{\tan \iota}{\sqrt{G M a\left(1-e^2\right)}} \frac{\partial \tilde{\mathcal{L}}_{\rm p}}{\partial \omega}, \\
        \dot{u} & =\sqrt{\frac{G M}{a^3}}-\sqrt{\frac{4 a}{G M}} \frac{\partial \tilde{\mathcal{L}}_{\rm p}}{\partial a}-\frac{1-e^2}{\sqrt{G M a e^2}} \frac{\partial \mathcal{L}_{\rm p}}{\partial e}, \\
        \dot{\omega} & =\sqrt{\frac{1-e^2}{G M a e^2}} \frac{\partial \mathcal{L}_{\rm p}}{\partial e}-\frac{\cos \iota}{\sqrt{G M a\left(1-e^2\right)} \sin \iota} \frac{\partial \mathcal{L}_{\rm p}}{\partial \iota}, \\
        \dot{\Omega} & =\frac{1}{\sqrt{G M a\left(1-e^2\right)} \sin \iota} \frac{\partial \mathcal{L}_{\rm p}}{\partial \iota},
    \end{align}
\end{subequations}
\endgroup
where $\mathcal{L}_{\rm p}$ is the Lagrangian corresponding to the perturbing force. The general force given in Eq.~\eqref{eq:Full force} implies a perturbing Lagrangian given by \citep{Kuntz_2023}
\begin{align}
    \mathcal{L}_{\rm p} = \frac{Q_\uf^{i j}}{2D}A_{1,ij} -Q_\uf^{i j} \frac{D_i D^k}{2D^3}A_{2,kj} + Q_\uf^{i j} \frac{D_i D_j D^k D^l}{4 D^5}A_{3,kl}.
\end{align}
where $A_{n,ij}$ are defined in Eq.~\eqref{eq:Full force}.

Note that post-Newtonian (PN) effects are not contained in the LPEs by default, but are typically incorporated by adding the corresponding PN terms to the perturbing Lagrangian, i.e. treating them as additional perturbative forces on the Newtonian orbit \citep{Kidder_1995,Mardling_2002}. While we accounted for PN effects in the near-field terms of the BG binary in Eq.~\eqref{eq:Full force}, but we do not add additional PN corrections here to the FG binary, which are generally subdominant%
\footnote{Note that in the early inspiral phase of the BG binary, the resonance condition together with the assumption of $M_\ub\gg M_\uf$ implies that $v_\uf/c = (M_\uf/M_\ub)^{1/3} v_\ub/c \ll v_\ub/c \ll 1$.},
but we wish to explore this in future work. In addition, we neglect the center-of-mass momentum kick imparted to the FG binary by the passing gravitational wave \citep[][Eq.~(32)]{Turner_1979}, as the resulting velocity change is entirely negligible and has neither theoretical nor observational consequences.

\paragraph{Maintaining Locking Across Zone Boundary---} 

Here, we demonstrate that the FG binary can remain locked as it crosses the near-zone/wave-zone boundary, i.e.~a FG binary system initially in resonance in the near-zone (where the dominant effect comes from the Newtonian quadrupole-quadrupole interaction) can remain locked as it moves into the wave-zone. This is shown in Figure~\ref{fig:example_locking_D=0.5}, where locking begins at $D_\lambdabar = 0.5$, and ends at $D_\lambdabar \approx 1.84$. However, if we are not careful, the strong $D^{-5}$ scaling of the Newtonian term can cause our 5 $n_\uf = n_\ub + k n_{\rm outer}$ resonances to overlap for small $D$, invalidating our treatment. Due to this, as well as our motivation to study GW-induced resonances, we will restrict ourselves to the wave-zone and leave near-zone resonances to future work.

\paragraph{Decoupling during Resonant Locking---}

\color{black} 
A binary system with a given $\iota_\uf$ and $\iota_{\rm outer}$ will decouple from resonance when there no longer exists a relative phase, $\Delta \phi_0$, such that Eq.~\eqref{eq:average change in FG frequency} is satisfied. We saw in Figure~\ref{fig:f_max_KL_contamination_T12345_Dlb=1} that such a $\Delta \phi_0$ exists in all $(\iota_{\rm f},\iota_{\rm outer})$ space when $D = \lambdabar$. Figure~\ref{fig:f_max_KL_contamination_T12345_Dlb=2} shows $\max(\overline{\dot{n_\uf}}/ \dot{n}_\ub)$ for $D = 2\lambdabar$. There are small regions of parameter space left that can sustain locking, but these also drop below unity by $D = 2.3\lambdabar$.

\color{black}

\paragraph{Galaxy Model---}
We select the galaxy's SMBH mass, which is the BG binary's mass $M_\ub$, and model our galaxy accordingly. Specifically, we focus on FG binaries in the galactic bulge. We assume the distribution of FG binaries is spherically symmetric. We assume that their distance and period distribution functions, given by $f_{D}(D)$ and $f_{T}(T)$ respectively, are independent. We assume the radial matter distribution in the bulge
\color{black} 
follows a Hernquist profile \cite{Hernquist_1990}
\begin{equation}
    f_{D}(D) \propto 4\pi D^2 \frac{1}{\left(\frac{D}{D_0}\right)\left(1 + \frac{D}{D_0}\right)^3} \propto 4\pi D^2 \rho_D(D),
\end{equation}
where $\rho_{D}(D)$ is the radial mass density distribution and $D_0 = R_e/1.8153$ is the bulge scale-length with $R_e$ being the half-light radius. When combining the $R_e$--$M_{\rm bulge}$ relation \citep{Hon_2022} with the $M_{\rm bulge}$--$M_{\rm SMBH}$ relation \citep[][Eq.~(10)]{Kormendy_2013}, one finds
\begin{align}
    D_0 &= 3.1312 \times \left(\frac{M_{\mathrm{b}}}{10^9 M_{\odot}}\right)^{0.748} {\rm kpc}.
\end{align}
\color{black}
Further, we normalize our distributions such that the total number of stars is $N_{\mathrm{stars}} = M_{\mathrm{bulge}}/\langle m_{\mathrm{star}} \rangle$ where \black{$\langle m_{\mathrm{star}} \rangle = 0.3 M_{\odot}$} is the average star mass \citep{Kroupa2002} and $M_{\mathrm{bulge}}$ comes from the $M_{\rm bulge}$--$M_{\rm SMBH}$ relation \citep[][Eq.~(10)]{Kormendy_2013}
\begin{align}
    M_{\mathrm{bulge}} &= 1.8 \times 10^{11} M_{\odot} \left(\frac{M_{\mathrm{b}}}{10^9 M_{\odot}}\right)^{0.85 \pm 0.06}. \label{eq:Mbulge MSMBH rel}
\end{align}
For the period distribution, we use Öpik's law \citep{Opik_1924} which states that $f_{T}(T)  = 1/(T\black{\ln\left[T_{\max}/T_{\min}\right]})$. \black{$T_{\max}$ is given in Equation \eqref{eq:Tmax B.C.}, while we choose $T_{\min}=10^{-5}$ days. Our choice for $T_{\min}$ is somewhat arbitrary, however the very weak logarithmic dependence ensures that its precise value has minimal impact on our results, even if changed by a few orders of magnitude.}

\color{black} 
\begin{figure*}[hbt!]
    \centering
        \subcaptionbox{$n_\uf = n_\ub + 2n_{\rm outer}, \iota_\uf = \pi, \iota_{\rm outer} = \pi$.
        \label{subfig:example_locking_+2Res.pdf}}
        {\includegraphics[width=.5\linewidth]{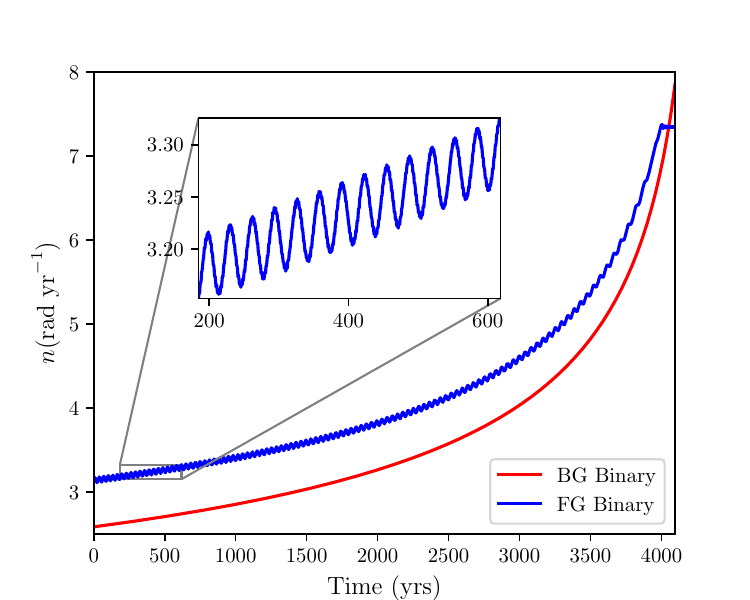}}\hfill 
        \subcaptionbox{$n_\uf = n_\ub + n_{\rm outer}, \iota_\uf = \pi/2, \iota_{\rm outer} = \pi/2$.
        \label{example_locking_+1Res.pdf}}
        {\includegraphics[width=.5\linewidth]{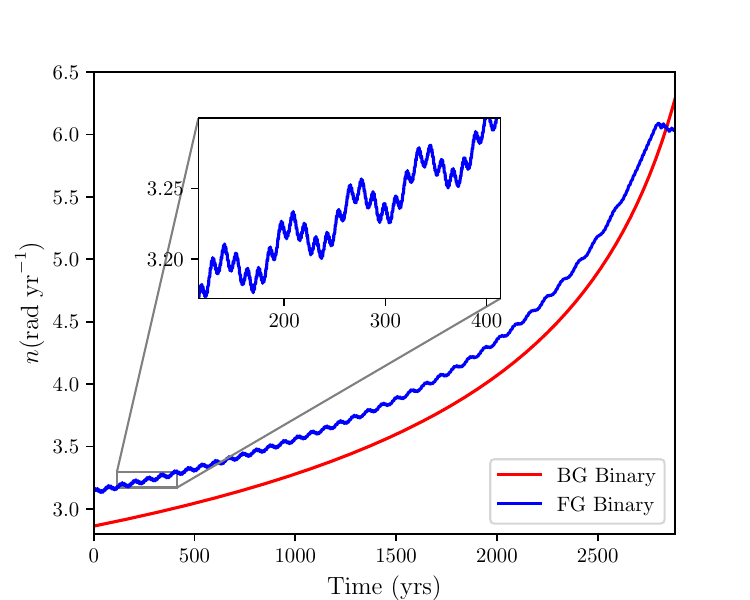}}\\
        \subcaptionbox{$n_\uf = n_\ub - n_{\rm outer}, \iota_\uf = \pi/2, \iota_{\rm outer} = 3\pi/2$.
        \label{subfig:example_locking_-1Res}}
        {\includegraphics[width=.5\linewidth]{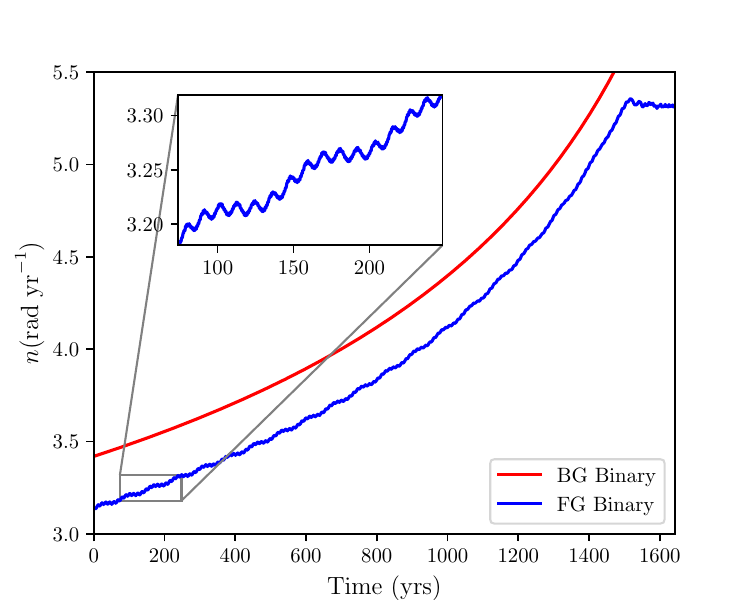}}\hfill 
        \subcaptionbox{$n_\uf = n_\ub - 2n_{\rm outer}, \iota_\uf = \pi, \iota_{\rm outer} = 0$.
        \label{subfig:example_locking_-2Res.pdf}}
        {\includegraphics[width=.5\linewidth]{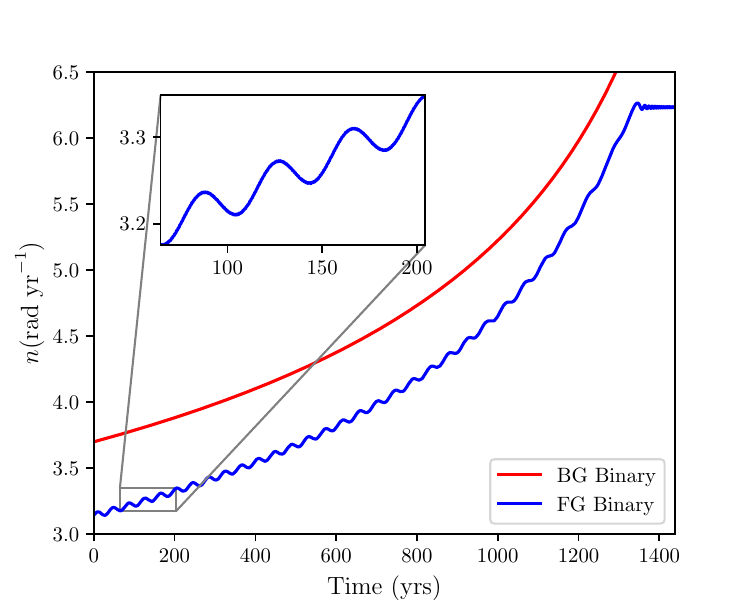}}\hfill
    \caption{\black{Numerically integrating Eq.~\eqref{eq:LPEs wrt M} for resonances $n_\uf = n_\ub + kn_{\rm outer}$ where $k = \{2,1,-1,-2\}$ in panels (a), (b), (c), (d), respectively. We have set $\mathcal{M}_\ub = 8.71 \times 10^8 M_{\odot}, \Dlb = 1, e_{\ub,0} = 0, \phi_{\rm b,0} = 0, T_{\rm b,0} = T_{\rm f,0} = 2 {\rm \ yrs}, e_{\uf,0} = 0.1$ initially, while all other unspecified orbital elements were initially set to zero.}}
    \label{fig:example_locking_diff_res}
\end{figure*}

\paragraph{Thrown into Resonance by Encounters---}
Here, we consider an encounter between the FG binary and a perturber of mass $M_{\mathrm{p}}$, impact parameter $b$, and velocity $v$, taken as the bulge velocity dispersion from the $M_{\rm SMBH}$–$\sigma$ relation (Eq.~\ref{M-sigma relation}). The interaction rate of such an encounter is 
\begin{align}
    \Gamma = \pi b^2 n_{\rm stars} v = \pi b^2 n_{\rm stars} \sigma \label{eq:interaction rate}
\end{align}
where $n_{\rm stars}$ is the local stellar number density. We use the separatrix treatment presented in our simplified toy model to give an estimate of the chance for the binary system to get kicked into the separatrix by a random scattering encounter by a third object. This puts upper and lower limits on the size of the kick. 

The separatrix height is (cf.~Figure~\ref{fig:Forced_pendulum_Separatrix}; this follows from the equation of motion, Eq.~\eqref{eq:gen EOM})
\begin{align}
    H_{\rm sep} = \sqrt{4 \sqrt{C_\uA^2 - C_\uF^2} + 4 C_\uF\sin^{-1}\left(\frac{C_\uF}{C_\uA}\right) - 2\pi C_\uF}.
\end{align}
We consider two cases: (1) the pendulum is close to exact resonance, namely on its final oscillation before reaching $\dot{\theta} = 0$, in which case we take the size of the kick to be such that $H_{\rm sep}/2 < \Delta n_\uf < 3H_{\rm sep}/2$, and (2) the pendulum is far from resonance, in which case $\left|n_\uf - n_\ub\right| - H_{\rm sep}/2 < |\Delta n_\uf| < \left|n_\uf - n_\ub\right| + H_{\rm sep}/2$.

We label the lower and upper bounds as $\Delta n_{\rm lower}$ and $\Delta n_{\rm upper}$, respectively. From Kepler's law  $3\Delta E_\uf/E_\uf = 2 \Delta n_\uf/n_\uf$ and hence, from Eq.~\eqref{eq:Delta E from encounter}, there exist impact parameters $b_{\rm min}$ and $b_{\rm max}$ such that $\Delta n_\uf$ sits within the desired range. However, there lies the simultaneous condition that $(v/b)T < \gamma$ for work to be done (see text below~Eq.~\eqref{M-sigma relation}), and thus for the binary's frequency to change. Thus, 
\begingroup
\allowdisplaybreaks
\begin{align}
    b_{\rm max} (t) &= \min \left[\left(\frac{6 G^2 M_{\mathrm{p}}^2}{\sigma^2 n_\uf^2 \left(\frac{|\Delta n_\uf|}{n_\uf}\right)_{\rm lower}}\right)^{1/4}, \sigma T_\uf/\gamma\right],\\
    b_{\rm min} (t) &= \min \left[\left(\frac{6 G^2 M_{\mathrm{p}}^2}{\sigma^2 n_\uf^2 \left(\frac{|\Delta n_\uf|}{n_\uf}\right)_{\rm upper}}\right)^{1/4}, \sigma T_\uf/\gamma\right].
\end{align} 
\endgroup
From Eq.~\eqref{eq:interaction rate}, the interaction rate of such encounters is
\begin{align}
    \Gamma_{\rm into} = \pi n_{\rm starts} \sigma \left(b_{\rm max}^2 - b_{\rm min}^2\right). \label{eq:gamma_into expression}
\end{align}

\paragraph{Thrown out of Resonance by Encounters---}
Again, continuing the toy separatrix model, getting kicked out of resonance would require an encounter where $|\Delta n_\uf| < H_{\rm sep}/2$. From this, we get a lower bound on the impact parameter of encounters that do not disrupt resonance
\begin{align}
    b_{\rm out} &= \min \left[\left(\frac{6 G^2 M_{\mathrm{p}}^2}{\sigma^2 n_\uf^2 \left(\frac{H_{\rm sep}}{2n_\uf}\right)}\right)^{1/4}, \sigma T_\uf/\gamma\right], \label{eq:b_out eqn}
\end{align}
where, from Eq.~\eqref{eq:interaction rate}, the corresponding interaction rate is 
\begin{align}
    \Gamma_{\rm out} = \pi n_{\rm starts} \sigma b^2_{\rm out}. \label{eq:gamma_out}
\end{align}

\paragraph{Resonant Locking for Different Resonances---}
Figure~\ref{fig:nf/nb for general treatment} shows $\max(\overline{\dot{n}}_\uf/\dot{n}_\ub)$ as a function of $\iota_{\rm f}$ and $\iota_{\rm outer}$ for all 5 resonances found in the general treatment, $n_\uf = n_\ub + kn_{\rm outer}$ for $k \in \{\pm 2, \pm 1\}$. Figure \ref{subfig:num_reses} shows the number of these resonances that have $\max(\overline{\dot{n}}_\uf/\dot{n}_\ub) > 1$ as a function of $\iota_{\rm f}$ and $\iota_{\rm outer}$.

As a means of demonstration, Figure~\ref{fig:example_locking_diff_res} shows resonant locking occurring for each of the 4 new resonances found in the general treatment, $n_\uf = n_\ub + kn_{\rm outer}$ for $k \in \{\pm 2, \pm 1\}$ ($k = 0$ is shown in Figure~\ref{fig:example_locking}). In all cases, the initial conditions are set such that the system starts within the separatrix analogous to the orange region in Figure~\ref{fig:Forced_pendulum_Separatrix}.

\paragraph{Integration Limits---}
As discussed above, we choose $T_{\rm min} = 10^{-5}$ days. $T_{\rm max}$ comes from the condition that the FG binary's separation distance must be less than each component's Hill radius and that it must not be so soft that it would evaporate within a Hubble time, $T_{\rm H}$, before the galaxy merger. The evaporation time of a binary with separation distance $a$ in an environment with velocity dispersion $\sigma$ and local mass density $\rho_D$ is \citep[][Eq.~(7.173)]{binney_2008}
\begin{align}
    t_{\rm evap} \sim \frac{0.061 \sigma}{G \rho_D a}. \label{eq:evap time}
\end{align}
We must have $t_{\rm evap} > T_{\rm H}$ for such a binary to still exist in our post-merger galaxy. Thus,
\begingroup
\allowdisplaybreaks
\begin{align}
    \begin{split}
        T_{\rm max}(D) &= \min \left[\frac{4\pi}{c} \sqrt{\frac{D^3}{6 r_{\rm b,s}}},\right.\\
        & \left. 2\pi \sqrt{\frac{1}{0.6 G M_{\odot}} \left(\frac{0.061 \sigma}{G \rho_D(D) T_{\rm H} \ln \Lambda}\right)^3}\right]. \label{eq:Tmax B.C.}
    \end{split}
\end{align}
\endgroup

\paragraph{Number of Locked Binaries over the History of All Mergers---}

This may be expressed as
\begin{align}
    N_{\rm ever} &= N_{\rm g}\int_{M_{\rm min}}^{M_{\rm max}} N_{\rm locked} \,\langle N_{\rm mergers} \rangle\, f_{\rm SMBH} \,\dd M_{\rm SMBH}\label{eq:N_ever}
\end{align}
where $N_{\rm g}$ is the number of galaxies in the observable Universe, $\langle N_{\rm mergers} \rangle\equiv\langle N_{\rm mergers} (M_{\rm SMBH}) \rangle$ is the average number of mergers a galaxy with SMBH mass $M_{\rm SMBH}$ experiences in its lifetime, and $f_{\rm SMBH}$ is the SMBH mass distribution function. We conservatively take $\langle N_{\rm mergers} \rangle = \mathcal{O}(1)$ to be unity. 

The SMBH mass distribution function is found from the DM halo mass function. Assuming there is one DM halo per SMBH, then using $f_{\rm SMBH} = f_{\rm DM} \ \dd M_{\rm DM}/\dd M_{\rm SMBH}$, with $f_{\rm DM} \propto M_{\rm DM}^{-1.806}$ \citep[][Figure~25a]{Kormendy_2013} (we never reach the exponential truncation of the mass function) and $\dd M_{\rm DM}/\dd M_{\rm SMBH}$ from Ref.~\citep[][Eq.~(18)]{Kormendy_2013}, we get
\begin{align}
    f_{\rm SMBH} = f_{\rm M,0}
    \begin{cases}
        M_{\rm SMBH}^{-1.298}, &\text{if } M_{\rm SMBH} < M_{\rm kink}\\
        \frac{M_{\rm kink}^{2.072}}{M_{\rm SMBH}^{3.370}}, &\text{if } M_{\rm SMBH} > M_{\rm kink}
    \end{cases}
\end{align}
where $M_{\rm kink} = 1.2 \times 10^8 M_{\odot}$ and $f_{\rm M,0}$ is the normalization constant. Assuming the smallest SMBH mass is $10^6 M_{\odot}$ and the largest is $10^{11} M_{\odot}$, we normalise the distribution and integrate Eq.~\eqref{eq:N_ever}. Assuming $N_{\rm g} = 2 \times 10^{12}$ \citep{Conselice_2016}, we find that
\begin{equation}
    N_{\rm ever} \approx 7.2 \times 10^{9}\,.
\end{equation} 

\end{document}